\documentclass[preprints,article,accept,moreauthors,pdftex]{Definitions/mdpi} 

\firstpage{1} 
\pubvolume{1}
\issuenum{1}
\articlenumber{0}
\pubyear{2021}
\copyrightyear{2021}
\externaleditor{Academic Editor: Ralph V. Chamberlin
} % For journal Automation, please change Academic Editor to "Communicated by"
\datereceived{29 April 2021} 
\dateaccepted{21 May 2021} 
\datepublished{} 
\hreflink{https://doi.org/} % If needed use \linebreak

\pdfoutput=1

\usepackage{bm}
\usepackage{braket}
\usepackage{changes}
\makeatletter
\@namedef{Changes@AuthorColor}{red}
\colorlet{Changes@Color}{red}
\makeatother

\Title{Quantum Heat Engines with Singular Interactions}

\TitleCitation{Quantum Heat Engines with Singular Interactions}

\Author{Nathan M. Myers $^{1,}$*\orcidA{}, Jacob McCready $^{1}$ and Sebastian Deffner $^{1,2,}$*\orcidB{}} 

\AuthorNames{Nathan M. Myers, Jacob McCready and Sebastian Deffner}

\AuthorCitation{Myers, N.M.; McCready, J.; Deffner, S.}
\address{%
$^{1}$ \quad Department of Physics, University of Maryland, Baltimore County, Baltimore, MD 21250, USA \\
$^{2}$ \quad Instituto de Física ‘Gleb Wataghin’, Universidade Estadual de Campinas, Campinas~13083-859,~São~Paulo,~Brazil}

\corres{Correspondence: myersn1@umbc.edu (N.M.M.); deffner@umbc.edu (S.D.)}

\abstract{By harnessing quantum phenomena, quantum devices have the potential to outperform their classical counterparts. Here, we examine using wave function symmetry as a resource to enhance the performance of a quantum Otto engine. Previous work has shown that a bosonic working medium can yield better performance than a fermionic medium. We expand upon this work by incorporating a singular interaction that allows the effective symmetry to be tuned between the bosonic and fermionic limits. In this framework, the particles can be treated as anyons subject to Haldane's generalized exclusion statistics. Solving the dynamics analytically using the framework of ``statistical anyons'', we explore the interplay between interparticle interactions and wave function symmetry on engine performance.}


\begin{document}
%%%%%%%%%%%%%%%%%%%%%%%%%%%%%%%%%%%%%%%%%%
\section{Introduction}

Thermodynamics was originally developed as a physical theory for the purpose of optimizing the performance of large-scale devices, namely steam engines \cite{Kondepudi1998}. Despite these practically-focused origins, thermodynamics has proven enormously successful in formulating universal statements, such as the second law. With the growing prominence of quantum technologies, the field of quantum thermodynamics has emerged to understand how the framework of thermodynamics can be extended to quantum systems \cite{Deffner2019book}. One of the principal goals of quantum thermodynamics is to discover how quantum features, such as entanglement, superposition, and coherence, can be best leveraged to optimize the performance of quantum devices. 

In the tradition of thermodynamics, the analysis of quantum heat engines has provided one of the primary tools for exploring how quantum effects change the thermodynamic behavior of a system. A non-exhaustive list of literature analyzing quantum thermal machine performance includes works examining the role of coherence \cite{Scully2003, Scully2011, Uzdin2016, Watanabe2017, Dann2020, Feldmann2012, Hardal2015, Hammam2021}, quantum correlations \cite{Barrios2021}, many-body effects \cite{Hardal2015, Beau2016, Li2018, Chen2019, Watanabe2020}, quantum uncertainty \cite{Kerremans2021}, relativistic effects \cite{Munoz2012, Pena2016, Papadatos2021}, degeneracy \cite{Pena2017, Barrios2018}, endoreversible cycles \cite{Deffner2018, Smith2020, Myers2021}, finite-time cycles \cite{Cavina2017, Feldmann2012, Zheng2016, Raja2020}, energy optimization \cite{Singh2020}, shortcuts to adiabaticity \cite{Abah2017, Abah2018, Abah2019, Beau2016, Campo2014, Funo2019, Bonanca2019, Baris2019, Dann2020, Li2018}, efficiency and power statistics \cite{Denzler2020, Denzler20202, Denzler20203}, and comparisons between classical and quantum machines \cite{Quan2007, Gardas2015, Friedenberger2017, Deffner2018}. Implementations have been proposed in a wide variety of systems including harmonically confined single ions \cite{Abah2012}, magnetic systems \cite{Pena2015}, atomic clouds \cite{Niedenzu2019}, transmon qubits \cite{Cherubim2019}, optomechanical systems \cite{Zhang2014, Dechant2015}, and quantum dots \cite{Pena2019, Pena2020}. Quantum heat engines have been {experimentally} implemented using nanobeam oscillators \cite{Klaers2017}, atomic collisions \cite{Bouton2021}, and two-level ions \cite{Horne2020}.      

A notable feature of quantum particles is that they are truly indistinguishable. To account for this, the wave function of a multiparticle state is constructed out of the symmetric (for bosons) or antisymmetric (for fermions) superposition of the single particle states. This wave function symmetry has physical consequences in terms of state occupancy, with any number of bosons allowed to occupy the same quantum state while fermions are restricted to single occupancy---the famous Pauli exclusion principle. The superposition arising from the symmetrization requirements also leads to interference effects that manifest as ``exchange forces" in the form of an effective attraction between bosons and effective repulsion between fermions \cite{Griffiths}. Wave function symmetry leads to quantum modifications of thermodynamic behavior, allowing for more work to be extracted from indistinguishable particles through mixing \cite{Yadin2021}, or as the working medium of a quantum heat engine~\cite{Jaramillo2016, Huang2017, Myers2020, Myers2021}. In ~\cite{Myers2020}, we showed that for a working medium of two non-interacting identical particles, a harmonic quantum Otto engine exhibits enhanced performance if the particles are bosons and reduced performance if they are fermions. In this paper we expand upon these results by introducing an interaction proportional to the inverse square of the interparticle distance in addition to the standard harmonic potential. This model is often referred to as the singular \cite{Nogueira2016} or isotonic oscillator \cite{Weissman1979}. 

This potential is of particular interest as it provides the basis for the Calogero--Sutherland model \cite{Calogero1969, Sutherland1988}, a system known to host generalized exclusion statistics (GES) anyons \cite{Haldane1991, Murthy1994}. In the framework of GES, the Pauli exclusion principle is generalized to allow for a continuum of state occupancy, from the single occupancy allowed to fermions up to the infinite state occupancy allowed to bosons \cite{Haldane1991}. In the Calogero--Sutherland model this anyonic behavior arises from tuning the interparticle interaction strength, effectively interpolating between the bosonic and fermionic exchange forces \cite{Murthy1994}.        
 
We determine a closed form for the thermal state density matrix for two particles in the singular oscillator potential. By mapping to an equivalent system of ``statistical anyons'' \cite{Myers2021} consisting of a statistical mixture of bosons and fermions, we determine the time-dependent internal energy while varying the oscillator potential. We then analyze a quantum Otto cycle, demonstrating performance that interpolates between the bosonic and fermionic of ~\cite{Myers2020} as the interaction strength parameter is changed.  

%%%%%%%%%%%%%%%%%%%%%%%%%%%%%%%%%%%%%%%%%%
\section{The Two-Particle Singular Oscillator}

We begin by outlining the model, notation, and previous results from the literature that are central to our analysis. We use the following Hamiltonian for two particles in a singular oscillator potential \cite{Ballhausen19882, Murthy1994},
\begin{equation}
	\label{2partHamil}
	H = \frac{\mathbf{p_1}^2+\mathbf{p_2}^2}{2m}+\frac{1}{2} m \omega^2 \left( \mathbf{x_1}^2+\mathbf{x_2}^2\right) + \frac{\hbar \nu (\nu-1)}{2 m (\mathbf{x_1}-\mathbf{x_2})^2},
\end{equation}
where the parameter $\nu$ quantifies the strength of the interparticle interaction. Following Calogero's original approach \cite{Calogero1969}, we rewrite this Hamiltonian in the center of mass and relative coordinate frame using the momentum coordinate transformations $\mathbf{P} = \mathbf{p_1}+\mathbf{p_1}$, $\mathbf{p} = (\mathbf{p_1}-\mathbf{p_1})/2$ and the position coordinate transformations $\mathbf{X} = (\mathbf{x_1}+\mathbf{x_1})/2$, $\mathbf{x} = \mathbf{x_1}-\mathbf{x_1}$. With this transformation, we can solve the dynamics of the center of mass and relative coordinates separately,
\begin{equation}
	\label{CMHamiltonian}
	H_{\mathrm{CM}} = \frac{\mathbf{P}^2}{2 M} + \frac{1}{2} M \omega^2 \mathbf{X}^2,
\end{equation}     
\begin{equation}
	\label{relHamiltonian}
	H_{\mathrm{rel}} = \frac{\mathbf{p}^2}{2 \mu} + \frac{1}{2} \mu \omega^2 \mathbf{x}^2 + \frac{\hbar \nu (\nu-1)}{4 \mu \mathbf{x}^2},
\end{equation}
where $M = 2m$ and $\mu = m/2$. We see that the center of mass Hamilton is identical to that of an unperturbed harmonic oscillator for a single particle of mass $2m$, while the relative motion Hamiltonian is identical to that of a singular oscillator with a single particle of mass $m/2$. Note that this approach is very commonly applied when solving the dynamics of classical interacting oscillator systems \cite{GoldsteinBook}.     

The eigenfunctions and eigenenergies of Equation~(\ref{CMHamiltonian}) are well known, and those of Equation~(\ref{relHamiltonian}) have been determined directly \cite{Calogero1969, Weissman1979}, using operator methods \cite{Ballhausen1988} and as a generalization of the Morse potential \cite{Nogueira2016}. The eigenfunctions are,
\begin{equation}
	\label{RelWF}
	\psi_{\mathrm{rel}}(x) = \left(\frac{\mu \omega}{\hbar}\right)^{\frac{1}{4}(1+2\nu)}\sqrt{\frac{n!}{\Gamma(n+\nu+1/2)}}x^{\nu}\mathrm{exp}\left(-\frac{\mu \omega x^2}{2 \hbar}\right) \mathrm{L}_n^{\nu-1/2}\left(\frac{\mu \omega}{\hbar}x^2\right),
\end{equation}
with the corresponding eigenenergies,
\begin{equation}
	\label{relEnergy}
	E_{\mathrm{rel}} = \hbar \omega \left(2n+\nu+\frac{1}{2}\right),
\end{equation}
where $n = 0,1,2,...$ and $\mathrm{L}_n^{\alpha}(x)$ is the generalized Laguerre polynomial \cite{AbramowitzBook}. 

For indistinguishable quantum particles, the total wave function must remain symmetric under particle exchange for bosons, and antisymmetric for fermions. In the center of mass and relative coordinate framework, the symmetry condition is satisfied by $\psi_{\mathrm{rel}}(x) = \psi_{\mathrm{rel}}(-x)$ for bosons and $\psi_{\mathrm{rel}}(x) = -\psi_{\mathrm{rel}}(-x)$ for fermions \cite{Sutherland1988}. Examining \mbox{Equation (\ref{RelWF})}, we see that the parity depends on the value of the interaction strength parameter $\nu$. Noting the relationship between the generalized Laguerre and Hermite \mbox{polynomials~\cite{Szego1939}},
\begin{equation}
	\begin{split}
	&\mathrm{H}_{2n}(x) = (-1)^n \, 2^{2n} \, n! \, \mathrm{L}_n^{-1/2}(x^2) \\
	&\mathrm{H}_{2n+1}(x) = (-1)^n \, 2^{2n+1} \, n! \, x \, \mathrm{L}_n^{1/2}(x^2), 
	\end{split}
\end{equation}
along with the Gamma function identity \cite{AbramowitzBook},
\begin{equation}
	\Gamma\left(n+\frac{1}{2}\right) = \frac{(2n)!}{4^n n!}\sqrt{\pi},
\end{equation}
we see that for $\nu = 0$, Equation~\eqref{RelWF} becomes, 
\begin{equation}
	\label{relBoson}
	\psi_{\mathrm{rel}}^{\nu = 0}(x)=\frac{1}{\sqrt{2^{2n} (2n)!}} \bigg(\frac{\mu \omega}{\pi \hbar} \bigg)^{1/4} e^{- \frac{\mu \omega x^2}{2 \hbar}} H_{2n} \bigg( \sqrt{\frac{\mu \omega}{\hbar}}x\bigg).
\end{equation}

Similarly, for $\nu = 1$, Equation~\eqref{RelWF} becomes,  
\begin{equation}
	\label{relFermion}
	\psi_{\mathrm{rel}}^{\nu = 1}(x)=\frac{1}{\sqrt{2^{2n+1} (2n+1)!}} \bigg(\frac{\mu \omega}{\pi \hbar} \bigg)^{1/4} e^{- \frac{\mu \omega x^2}{2 \hbar}} H_{2n+1} \bigg( \sqrt{\frac{\mu \omega}{\hbar}}x\bigg).
\end{equation}

{For $\nu = 0$ and $\nu = 1$, we see that the interaction potential vanishes, reducing \mbox{Equation (\ref{2partHamil})} to the Hamiltonian of two particles in a pure harmonic potential. We note that Equations (\ref{relBoson}) and (\ref{relFermion}) correspond to the eigenfunctions of the relative motion Hamiltonian for two bosons and two fermions in a pure harmonic potential, respectively.} The restriction of Equation (\ref{relBoson}) to even and Equation (\ref{relFermion}) to odd Hermite polynomials ensures that the proper exchange symmetry conditions are satisfied. This demonstrates that interacting particles in the singular oscillator potential can be treated as noninteracting anyons in a harmonic potential obeying generalized exclusion statistics \cite{Haldane1991}, with $\nu$ as the parameter that controls the nature of the particle statistics. This behavior was first established by Murthy and Shankar in the context of the thermodynamics of the Calogero--Sutherland model \cite{Murthy1994}. 

\section{Singular Oscillator Thermal State}

The equilibrium thermodynamic behavior of the two-particle singular oscillator can be determined from the canonical partition function,
\begin{equation}
	Z = \mathrm{tr}\{\mathrm{exp}(-\beta H)\}. 
\end{equation}

Noting $H = H_{\mathrm{CM}}+H_{\mathrm{rel}}$, the partition function can be split into the product of individual partition functions for the center of mass and relative motion. The center of mass partition function can be found straightforwardly in the energy representation using the harmonic oscillator eigenenergies,
\begin{equation}
	Z_{\mathrm{CM}} = \sum_{N = 0}^{\infty}\exp(-\beta E_{\mathrm{CM}}) = \sum_{N = 0}^{\infty} \exp\left(-\beta \hbar \omega \left[N+\frac{1}{2}\right]\right) = \frac{\exp(\beta \hbar \omega/2)}{\exp(\beta \hbar \omega)-1}. 
\end{equation}

We can find the relative partition function using Equation~(\ref{relEnergy}),
\begin{equation}
	Z_{\mathrm{rel}} = \sum_{n = 0}^{\infty}\exp(-\beta E_{\mathrm{rel}}) = \sum_{n = 0}^{\infty} \exp\left(-\beta \hbar \omega \left[2n+\nu+\frac{1}{2}\right]\right) = \frac{\exp\left(\beta \hbar \omega[3/2- \nu ] \right)}{\exp\left(2 \beta \hbar  \omega  \right)-1}. 
\end{equation}

The total canonical partition function is then, 
\begin{equation}
	\label{partFcn}
	Z = Z_{\mathrm{CM}}Z_{\mathrm{rel}} = \frac{\mathrm{exp}(-\beta \hbar \omega [\nu-2])}{[\mathrm{exp}(\beta \hbar \omega)+1][\mathrm{exp}(\beta \hbar \omega)-1]^2} 
\end{equation}

Equation~\eqref{partFcn} can be identified as a product of bosonic and fermionic contributions.  To this end, note that the partition functions for two bosons and two fermions in a harmonic potential are \cite{Myers2021},

\vspace{-3pt}
%\begin{adjustwidth}{-4.6cm}{0cm}
\begin{equation}
	Z_{\mathrm{B}} = \frac{\mathrm{exp}(2 \beta \hbar \omega)}{[\mathrm{exp}(\beta \hbar \omega)+1][\mathrm{exp}(\beta \hbar \omega)-1]^2},
\end{equation}
and,
\begin{equation}
	Z_{\mathrm{F}} = \frac{\mathrm{exp}(\beta \hbar \omega)}{[\mathrm{exp}(\beta \hbar \omega)+1][\mathrm{exp}(\beta \hbar \omega)-1]^2}.
\end{equation}
%\end{adjustwidth} 
Therefore we can express the singular oscillator partition function \eqref{partFcn} as,
\begin{equation}
	Z = \left(Z_{\mathrm{B}}\right)^{1-\nu}\left(Z_{\mathrm{F}}\right)^{\nu}.
\end{equation}

This agrees with the partition function determined by Murthy and Shankar for the Calogero--Sutherland model \cite{Murthy1994} and that of a statistical mixture of bosons and fermions in a harmonic potential, using the framework of statistical anyons \cite{Myers2021}. 

A fuller thermodynamic picture arises from the equilibrium thermal density matrix. This can be determined in position representation using,
\begin{equation}
	\label{DenForm}
	\rho(x,y) = \sum_{n = 0}^{\infty} \frac{1}{Z} \exp(-\beta E_n)\psi_n^*(x)\psi_n(y).
\end{equation}

The density matrix for the center of mass motion is given by the known thermal state of the quantum harmonic oscillator \cite{Greiner1995},
\begin{equation}
\label{eq:CM}
	\begin{split}
	\rho_{\mathrm{CM}}(X,Y) = & \frac{1}{Z_{\mathrm{CM}}} \sqrt{\frac{M \omega  \text{csch}(\beta \hbar \omega)}{2 \pi \hbar}} \\
	& \quad \times \exp \left(-\frac{M \omega}{4 \hbar} \left[(X+Y)^2 \tanh \left(\frac{\beta \hbar \omega}{2}\right)+(X-Y)^2 \coth \left(\frac{\beta \hbar \omega}{2}\right)\right]\right).
	\end{split}
\end{equation}

The density matrix for the relative motion can be found in closed form by combining Equations~(\ref{RelWF}) and (\ref{DenForm}) and applying the Hardy–Hille formula \cite{Bateman1953},
\begin{equation}
	\sum_{n = 0}^{\infty} \frac{n!}{\Gamma(n+\alpha+1)}\mathrm{L}_n^{\alpha}(x)\mathrm{L}_n^{\alpha}(y)t^n = \frac{1}{(xyt)^{\alpha/2}(1-t)}\exp\left(-\frac{(x+y)t}{1-t}\right)\mathrm{I}_{\alpha}\left(\frac{2\sqrt{xyt}}{1-t}\right), 
\end{equation}
where $\mathrm{I}_{\alpha}(x)$ is the modified Bessel function of the first kind \cite{AbramowitzBook}. This yields,
\begin{equation}
\label{eq:rel}
	\begin{split}
	\rho_{\mathrm{rel}}(x,y) = & \frac{1}{Z_{\mathrm{rel}}}\frac{\mu  \omega}{2\hbar}  \text{csch}(\beta \hbar \omega) \sqrt{x y} \\
	& \quad \times \exp\left(-\frac{\mu  \omega}{2 \hbar }  \left(x^2+y^2\right) \coth (\beta \hbar \omega)\right) \mathrm{I}_{\nu -\frac{1}{2}}\left(\frac{\mu \omega}{\hbar} x y \, \text{csch}(\beta \hbar \omega)\right).
	\end{split}
\end{equation}

The total thermal state density matrix in the position representation can then be determined from the simple Cartesian product of $\rho_{\mathrm{CM}}$ and $\rho_{\mathrm{rel}}$, Equations~\eqref{eq:CM} and \eqref{eq:rel}, respectively, 
\begin{equation}
\rho(X,Y,x,y) = \rho_{\mathrm{CM}}(X,Y) \rho_{\mathrm{rel}}(x,y).
\end{equation}                                 

\section{Singular Oscillator Dynamics}

With the equilibrium behavior of the two-particle singular oscillator established, we next consider the evolution of the system for a time-dependent parameterization of the frequency, $\omega = \omega(t)$. As in the case of the thermal state, we will treat the dynamics of the relative and center of mass motion separately.     

The dynamics of the center of mass coordinate, described by Equation~(\ref{CMHamiltonian}), are that of the well-studied time-dependent parametric harmonic oscillator \cite{Husimi1953}. The time-evolved state can be determined using the path integral formulation by applying the appropriate propagator,
\begin{equation}
	\label{EvolvedCM}
	\rho_{\mathrm{CM}}(X,Y;t) = \int \mathrm{d}X_0 \int \mathrm{d}Y_0 \, U_{\mathrm{CM}}(X,X_0;t) \, \rho_{\mathrm{CM}}(X_0,Y_0;0) \, U_{\mathrm{CM}}^{\dagger}(Y,Y_0;t), 
\end{equation}
where $U_{\mathrm{CM}}(X,X_0;t)$ is given by \cite{Husimi1953},
\begin{equation}
	U_{\mathrm{CM}}(X,X_0;t) = \sqrt{\frac{M}{2 \pi i h Z_t}}\exp\left(\frac{iM}{2\hbar Z_t} (\dot{Z_t}X^2-2XX_0+W_tX_0^2)\right).
\end{equation} 
$Z_t$ and $W_t$ are time-dependent solutions to the classical harmonic oscillator equation of motion,
\begin{equation}
	\label{ClassicalEOM}
	\ddot{Z}_t + \omega_t^2 Z_t = 0,
\end{equation}
with initial conditions $Z_0 = 0$, $\dot{Z}_0 = 1$ and $W_0 = 1$, $\dot{W}_0 = 0$.

The dynamics of the relative coordinate, described by Equation~(\ref{relHamiltonian}), are that of the time-dependent parametric singular oscillator. The corresponding propagator for this Hamiltonian has also been determined exactly \cite{Dodonov1972, Dodonov1975, Khandekar1986, Dodonov1992},   

%\begin{adjustwidth}{-4.6cm}{0cm}
\begin{equation}
	\begin{split}
	U_{\mathrm{rel}}(x,x_0,t) = &\frac{m \sqrt{x x_0}}{\sqrt{2}\hbar Z_t} \, \mathrm{J}_{\nu-1/2}\left(\frac{m x x_0}{\hbar Z_t}\right) \\ & \qquad \times \mathrm{exp}\left(-\frac{1}{2}i(\nu+1/2)\pi+\frac{i m}{2 \hbar Z_t}(\dot{Z_t}x^2+W_t x_0^2)\right),
	\end{split}
\end{equation}
%\end{adjustwidth}
where $\mathrm{J}_{\nu}(x)$ denotes the Bessel function of the first kind \cite{AbramowitzBook} and $Z_t$ and $W_t$ are the same as in Equation~(\ref{ClassicalEOM}). As in the center of mass case, the time-evolved density matrix for the relative motion is given by,
\begin{equation}
	\label{Evolvedrel}
	\rho_{\mathrm{rel}}(X,Y;t) = \int \mathrm{d}x_0 \int \mathrm{d}y_0 \, U_{\mathrm{rel}}(x,x_0;t) \, \rho_{\mathrm{rel}}(x_0,y_0;0) \, U_{\mathrm{rel}}^{\dagger}(y,y_0;t).
\end{equation}

While the integrals in Equation~(\ref{EvolvedCM}) can be carried out analytically \cite{Deffner2008,Deffner2010CP,Deffner2013PRE}, the product of Bessel functions in Equation~(\ref{Evolvedrel}) makes determining a closed-form expression for general values of $\nu$ difficult. To determine an analytical expression for the time-evolved state, we instead turn to the framework of statistical anyons \cite{Myers2021}.

In \cite{Myers2021}, we showed that the behavior of a pair of particles in a non-interacting statistical mixture of boson and fermion pairs is, on average, fully equivalent to that of interacting particles obeying generalized exclusion statistics. In this framework, the generalized exclusion statistics parameter is equivalent to the probability that a given pair of particles in the statistical mixture are fermions, $p_{\mathrm{F}}$ \cite{Myers2021}. In the context of this work, this means that we can exactly map the behavior of particles in a singular oscillator potential to the average behavior of a statistical mixture of bosons and fermions in a \textit{harmonic} oscillator potential. Using this mapping, the singular oscillator density matrix is given by,
\begin{equation}
	\label{SArho}
	\rho(x_1,x_2,y_1,y_2;t) = (1-p_{\mathrm{F}}) \, \rho_{\mathrm{har}}^{(\mathrm{B})}(x_1,x_2,y_1,y_2;t) + p_{\mathrm{F}} \, \rho_{\mathrm{har}}^{(\mathrm{F})}(x_1,x_2,y_1,y_2;t),
\end{equation}
where $\rho_{\mathrm{har}}^{(\mathrm{B})}(x_1,x_2,y_1,y_2;t)$ and $\rho_{\mathrm{har}}^{(\mathrm{F})}(x_1,x_2,y_1,y_2;t)$ are the time-evolved states for two non-interacting bosons and fermions in a harmonic potential, respectively. The full expressions were determined in \cite{Myers2020} and are provided in Appendix \ref{AppendixA} for completeness. Note that in Equation~(\ref{EvolvedCM}), we have switched back into the individual particle frame. Recalling that the generalized exclusion statistics parameter for the singular oscillator is given by the interaction strength \cite{Murthy1994}, we have the relation $p_{\mathrm{F}} = \nu$. 

The time-dependent average energy can be determined in the usual manner,
\begin{equation}
	\langle H_t \rangle = \mathrm{tr}\left\{\rho(x_1,x_2,y_1,y_2;t)H_t \right\}. 
\end{equation}

Using Equation~(\ref{SArho}), this becomes,
\begin{equation}
	\label{HSA}
	\langle H_t \rangle = (1-\nu)\langle H_{\mathrm{har}}^{(\mathrm{B})} \rangle + \nu\langle H_{\mathrm{har}}^{(\mathrm{F})} \rangle. 
\end{equation}

In \cite{Myers2020}, we found $\langle H_{\mathrm{har}}^{(\mathrm{B})} \rangle$ and $\langle H_{\mathrm{har}}^{(\mathrm{F})} \rangle$ to be,
\begin{equation}
	\label{HB}
	\langle H_{\mathrm{har}}^{(\mathrm{B})} \rangle = \frac{\hbar \omega_t}{2}\, Q^*\, \left( 3 \mathrm{coth}(\beta \hbar \omega_0) + \mathrm{csch}(\beta \hbar \omega_0) - 1 \right),
\end{equation}
and,
\begin{equation}
	\label{HF}
	\langle H_{\mathrm{har}}^{(\mathrm{F})} \rangle = \frac{\hbar \omega_t}{2}\, Q^*\, \left( 3 \mathrm{coth}(\beta \hbar \omega_0) + \mathrm{csch}(\beta \hbar \omega_0) + 1 \right),
\end{equation}
where {$\omega_0$ is the frequency at $t = 0$ and} $Q^*$ is a dimensionless parameter that measures the degree of adiabaticity of the evolution \cite{Husimi1953},
\begin{equation}
	Q^*= \frac{1}{2 \omega_0 \omega_t}\left[\omega^{2}_0\left(\omega^{2}_t Z_t^2 + \dot{Z}_t^2\right)+\left(\omega^{2}_t W_t^2+\dot{W}_t^2\right)\right].
\end{equation} 

For a perfectly adiabatic stroke, $Q^* = 1$, and in general, $Q^* \geq 1$ \cite{Husimi1953, Abah2012}. By combining Equations (\ref{HSA})--(\ref{HF}), we find,
\begin{equation}
	\label{tdependE}
	\langle H_t \rangle = \frac{\hbar  \omega_t}{2} Q^* \left(3 \coth \left(\beta  \omega_0 \hbar \right)+\text{csch}\left(\beta  \omega_0 \hbar \right)+2 \nu -1\right).
\end{equation}

Note that $\langle H_t \rangle = (\omega_t/\omega_0)Q^* \langle H_0 \rangle$, where $\langle H_0 \rangle$ is the internal energy of the equilibrium thermal state, which can be determined from Equation (\ref{partFcn}) using,
\begin{equation}
	\langle H_0 \rangle = - \frac{\partial }{\partial \beta} \ln(Z).
\end{equation}

This expression for $\langle H_t \rangle$ agrees with the result of \cite{Jaramillo2016}, which determined the time-dependent average energy of the singular oscillator using the scale invariance of the potential.                       
          
\section{The Quantum Otto Cycle}

\begin{figure}[H]
	\includegraphics[width=9 cm]{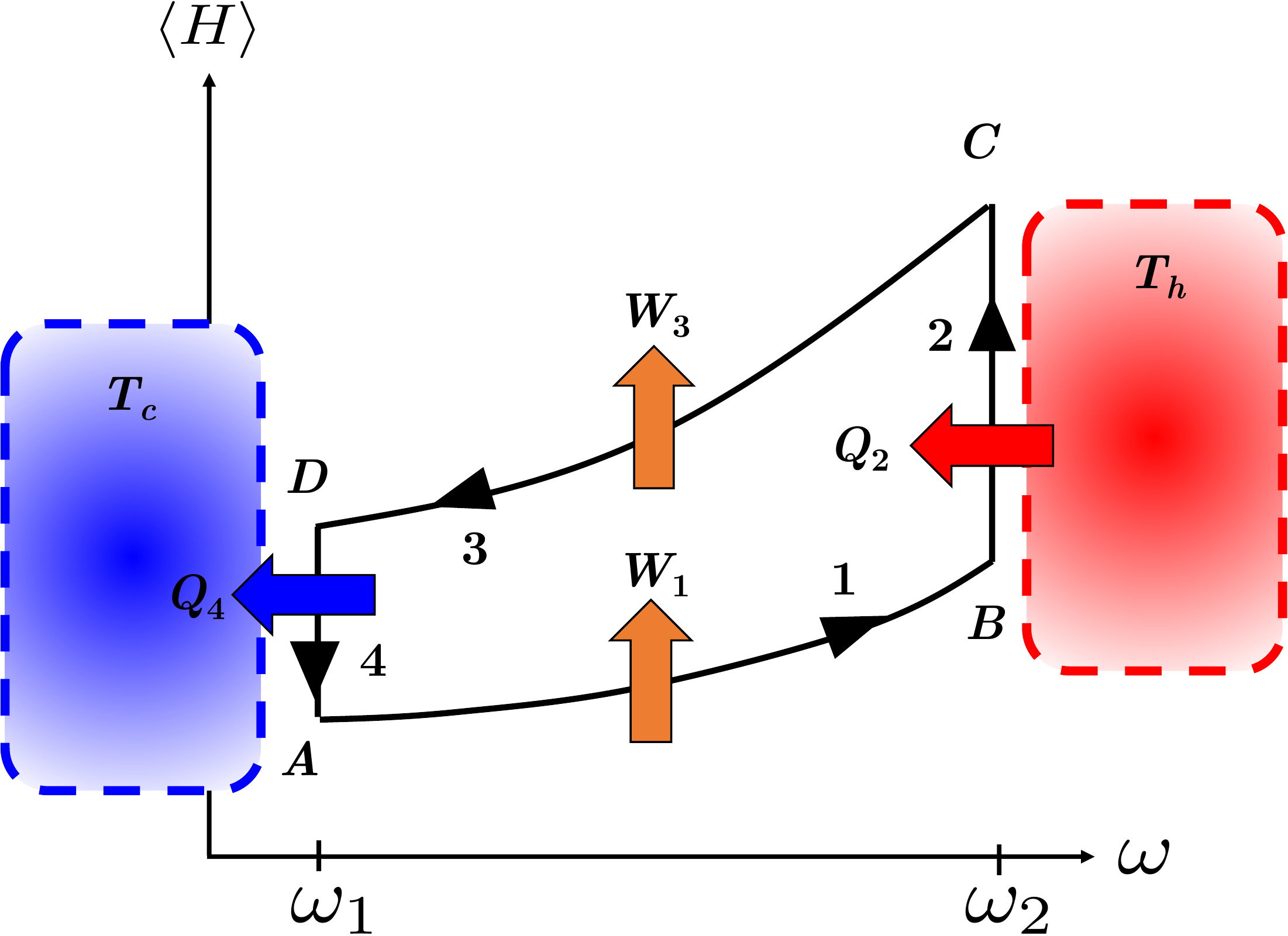}
	\caption{\label{cycleFig} Energy--frequency diagram of a quantum Otto cycle.}
\end{figure}

The quantum Otto cycle, like its classical counterpart, consists of four strokes: (1) isentropic compression, (2) isochoric heating, (3) isentropic expansion, and (4) isochoric cooling. For our working medium of two particles confined within a singular oscillator, the isentropic strokes consist of increasing the oscillator frequency for compression and decreasing it for expansion. {The cycle is illustrated graphically in Figure~\ref{cycleFig}.} During an isentropic stroke, the state of the working medium evolves unitarily with constant von Neumann entropy, thus fulfilling the isentropic condition. This is notably different from the classical Otto cycle in which the isentropic strokes are thermodynamically adiabatic and reversible. For the quantum cycle, the finite time unitary strokes are not generally adiabatic (i.e., in accordance with the quantum adiabatic theorem) and thus subject to ``quantum friction'' in the form of nonadiabatic excitations to other energy states \cite{Deffner2019book}.  

The isochoric strokes consist of coupling the working medium to a hot (cold) bath to increase (decrease) the energy of the system while holding the oscillator frequency constant. We make the standard assumption that the thermalization time is sufficiently short such that the working medium has achieved a state of thermal equilibrium with the bath by the end of each isochoric strokes, which removes the need to explicitly model the system--bath interaction \cite{Kosloff1984, Rezek2006, Abah2012, Campo2014, Beau2016, Abah2017, Myers2020, Watanabe2020, Myers2021}.  {Note that, unlike the classical quasistatic Otto cycle, the quantum Otto cycle is fundamentally irreversible \cite{Deffner2019book}. The working medium is only in a state of thermal equilibrium with the cold and hot baths at points $A$ and $C$, respectively. The nonadiabatic nature of the isentropic strokes will drive the system away from these equilibrium states during the rest of the cycle. The thermalization of the resulting out-of-equilibrium states during the isochoric strokes is then the source of the irreversibility of the cycle \cite{Deffner2019book}.}

We consider a linear driving protocol for duration $\tau$, such that the time dependence of the frequency is given by,
\begin{equation}
	\omega_t = \left( \omega_0^2 + \delta\omega \frac{t}{\tau}\right)^{1/2}, 
\end{equation}
for the compression stroke and, 
\begin{equation}
	\omega_t = \left( \omega_{\tau}^2 - \delta\omega \frac{t}{\tau}\right)^{1/2}, 
\end{equation}
for the expansion stroke. Note that for this driving protocol, Equation~(\ref{ClassicalEOM}) can be solved analytically and yields solutions in terms of the Airy functions \cite{Deffner2008}.  

To determine the average heat and work exchanged during each stroke, we use the following method. We determine the internal energy at the beginning of the compression stroke (point $A$ in the cycle) from a thermal equilibrium state with the cold bath. Then using Equation~(\ref{tdependE}), we determine the internal energy at the end of the compression stroke (point $B$). The change in internal energy between $A$ and $B$ gives the average work done on the system ($W_1$). We can find the the internal energy at point $C$ using the thermal equilibrium state with the hot bath. The change in internal energy between $C$ and $B$ gives the average heat exchanged with the hot bath ($Q_2$). Again, applying Equation~(\ref{tdependE}), we find the internal energy at the end of the expansion stroke (point $D$). The change in internal energy between $C$ and $D$ gives the average work done by the system ($W_2$). Finally, we can then use the difference in internal energy between points $D$ and $A$ to find the average heat exchanged with the cold bath ($Q_4$).      

With the average work and heat for each stroke in hand, we can then calculate the efficiency from the ratio of the total work to the heat input and the power from the ratio of total work to the cycle time,    
\begin{equation}
	\eta = -\frac{\langle W_1 \rangle + \langle W_3 \rangle}{\langle Q_2 \rangle} \quad \text{and}\quad P = -\frac{\langle W_1 \rangle + \langle W_3 \rangle}{\tau_{\mathrm{cyc}}}\,.
\end{equation}

We consider isentropic strokes of equal duration, $\tau_1 = \tau_3 = \tau$. As we do not explicitly model the system--bath interaction, we represent the duration of the isochoric strokes as a multiplicative factor of the isentropic stroke duration. Thus, we can represent the total cycle time as $\tau_{\mathrm{cyc}} = 2\gamma\tau$. The full expressions for the efficiency and power are cumbersome and detailed in Appendix \ref{AppendixB}. 

\begin{figure}[H]
	\includegraphics[width=10 cm]{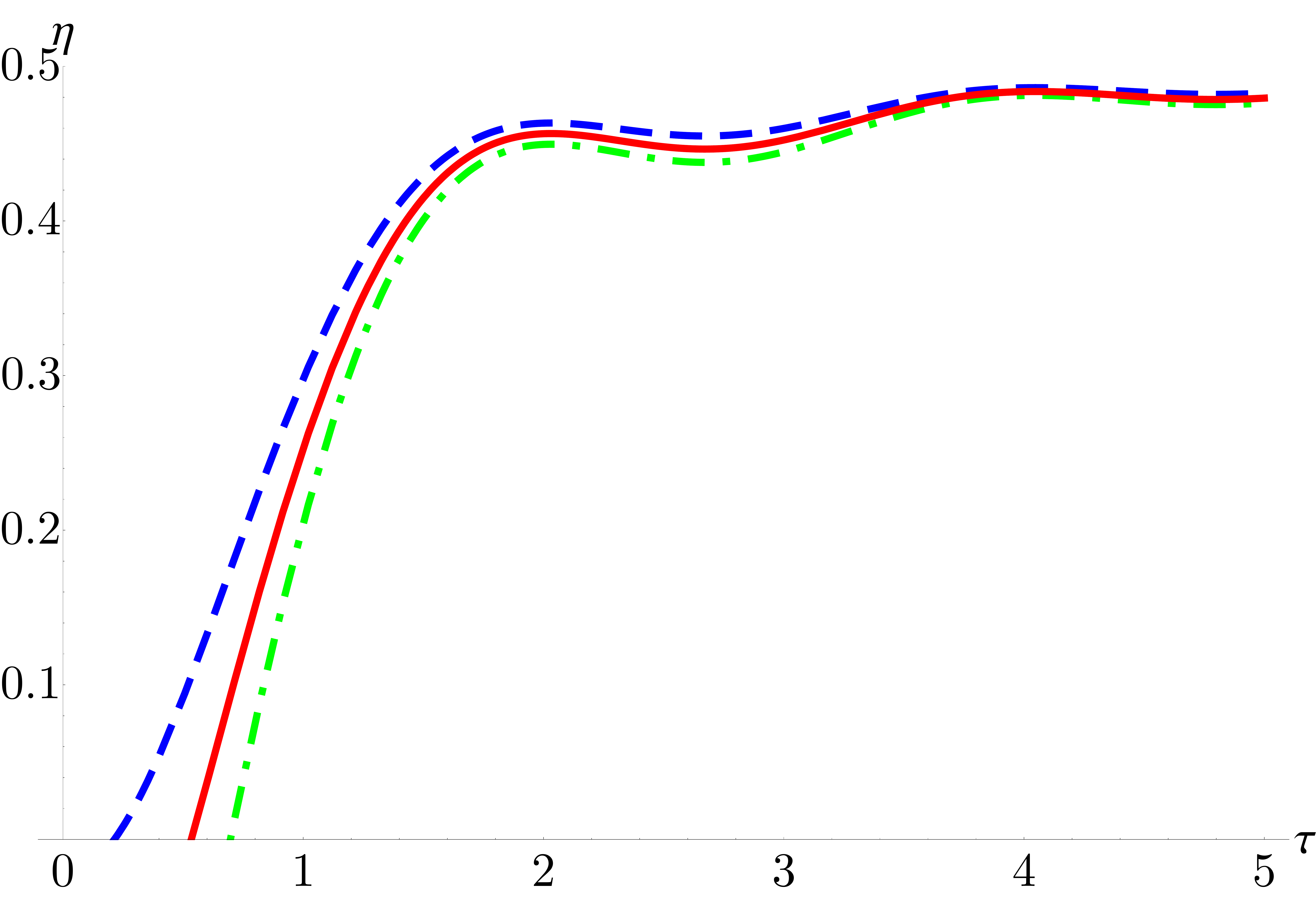}
	\caption{\label{effvTauFig} Efficiency as a function of the stroke time for $\nu = 0$ (blue, dashed), $\nu = 1/2$ (red, solid), and $\nu = 1$ (green, dot-dashed). Parameters are $\omega_0 = T_{\mathrm{c}} = 1, T_{\mathrm{h}} = 4, \delta\omega = 3$.}
\end{figure}

In Figure~\ref{effvTFig}, we plot the engine efficiency as a function of the ratio of bath temperatures. Confirming the results of \cite{Myers2020}, we see that the efficiency is greatest in the non-interacting bosonic limit of $\nu = 0$ and least in the non-interacting fermionic limit of $\nu = 1$. Between these limits, we observe that increasing the interaction strength from zero to one interpolates the efficiency smoothly between the bosonic and fermionic bounds. In Figure~\ref{effvTauFig}, we plot the engine efficiency as a function of the stroke time, $\tau$. Again, we see that $\nu = 0$ provides the greatest efficiency and $\nu = 1$ the worst, with intermediate values of $\nu$ falling between these limits. We see that increasing the stroke time results in increasing efficiency, approaching the bound of $\eta = 1 - \omega_0/\omega_{\tau}$ achieved in the limit of perfectly adiabatic strokes ($Q^*=1$). We note that in the limit of long stroke times the efficiency converges to this limit for all values of $\nu$. This indicates that the influence of the interaction on the engine performance is a fundamentally nonequilibrium effect. The oscillatory behavior of the efficiency arises from the form of $Q^*$ for the linear protocol.     

\begin{figure}[H]
	\includegraphics[width=9 cm]{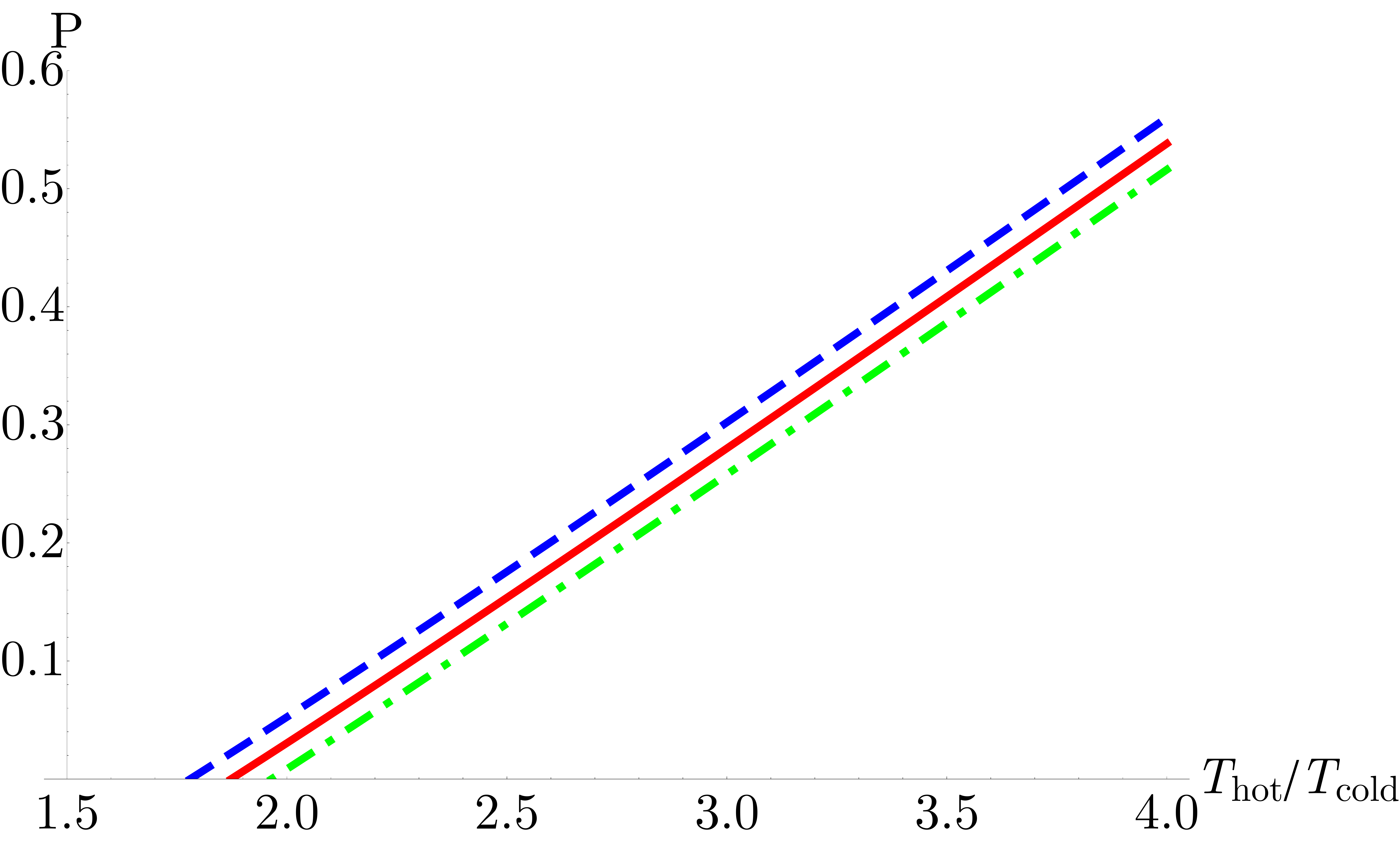}
	\caption{\label{PvTFig} Power as a function of the bath temperature ratio for $\nu = 0$ (blue, dashed), $\nu = 1/2$ (red, solid), and $\nu = 1$ (green, dot-dashed). Parameters are $\omega_0 = \delta\omega = \tau = 1$.}
\end{figure}

\begin{figure}[H]
	\includegraphics[width=9 cm]{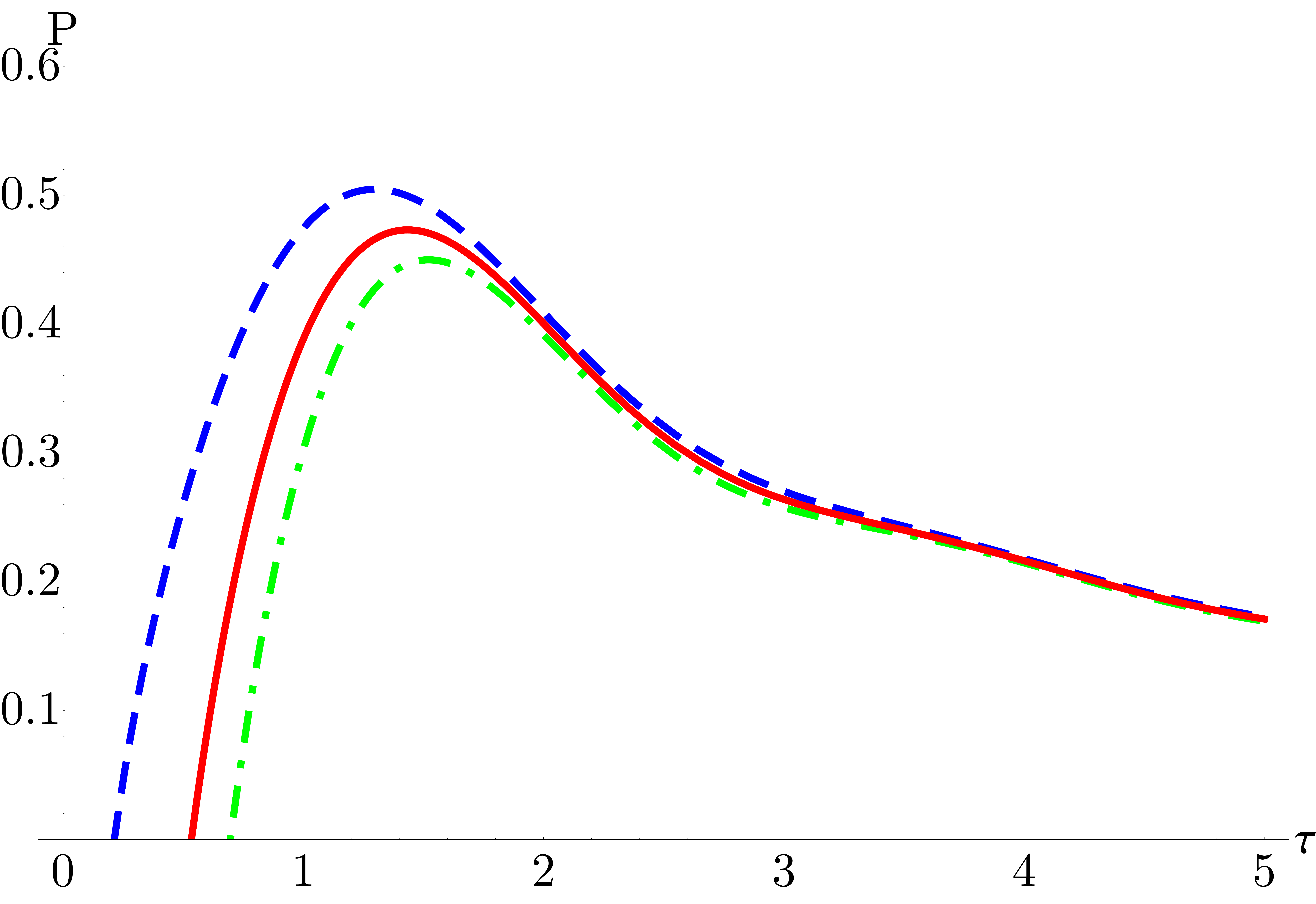}
	\caption{\label{PvTauFig} Power as a function of the stroke time for $\nu = 0$ (blue, dashed), $\nu = 1/2$ (red, solid), and $\nu = 1$ (green, dot-dashed). Parameters are $\omega_0 = T_{\mathrm{c}} = 1, T_{\mathrm{h}} = 4, \delta\omega = 3$.}
\end{figure}

Next, we examine the power output of the engine as a function of the ratio of bath temperatures, as shown in Figure~\ref{PvTFig}. As with the efficiency, we see that the bosonic limit ($\nu = 0$) gives the greatest power output, and the fermionic limit ($\nu = 1$) gives the least. Intermediate values of $\nu$ fall between these bounds. In Figure~\ref{PvTauFig}, we plot the power as a function of the stroke time. As in the case of the efficiency, we see that the shift in the power output arising from the interaction vanishes as the stroke time increases.

\begin{figure}[H]
	\includegraphics[width=10 cm]{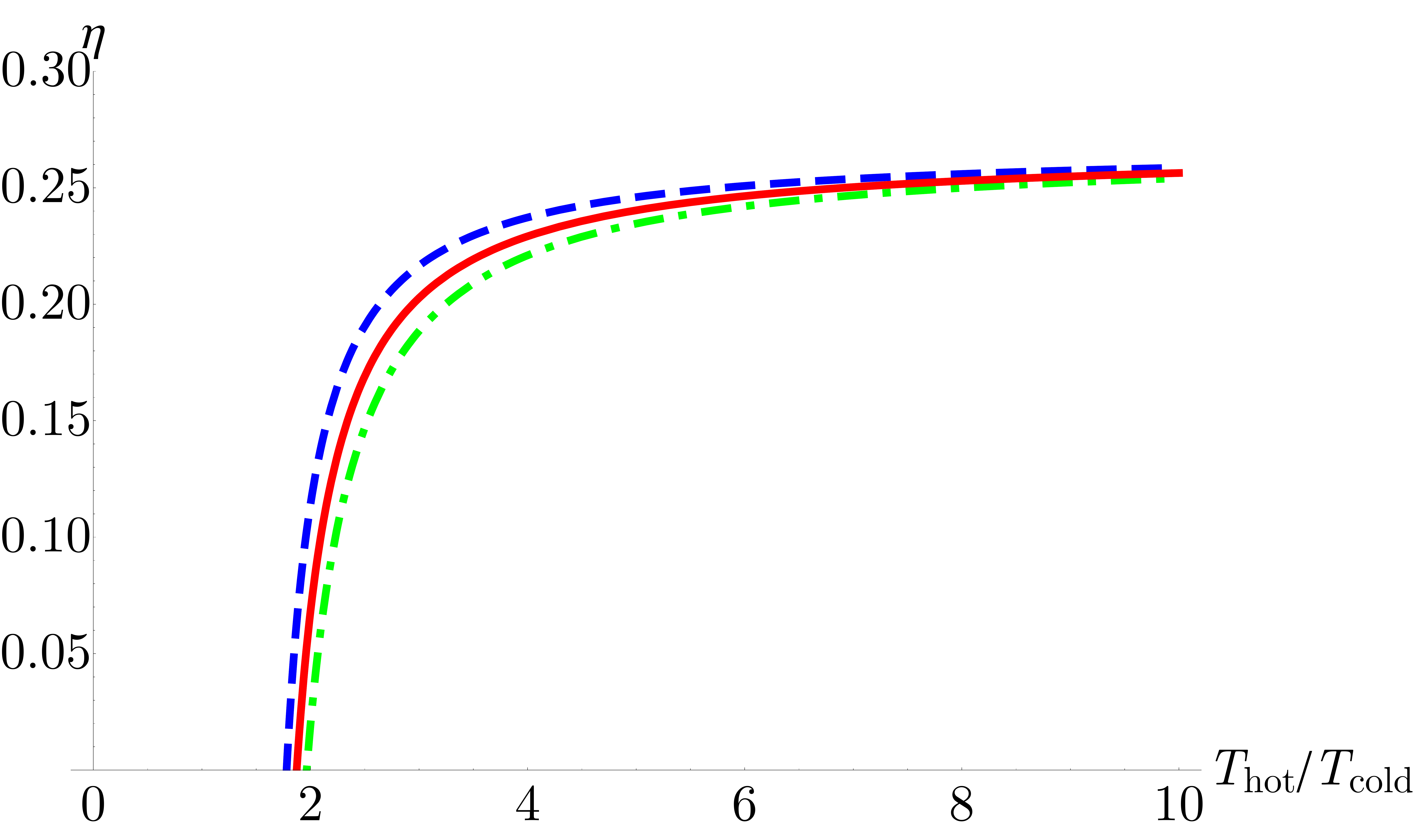}
	\caption{\label{effvTFig} Efficiency as a function of the bath temperature ratio for $\nu = 0$ (blue, dashed), $\nu = 1/2$ (red, solid), and $\nu = 1$ (green, dot-dashed). Parameters are $\omega_0 = \delta\omega = \tau = 1$.}
\end{figure} 

\vspace{-6pt}

\vspace{-6pt}

\vspace{-6pt}

Examining both efficiency and power plots, we observe that the zeroes for efficiency and power all occur at finite values of the bath temperature ratio or stroke time. These zeroes mark the transition to the parameter regimes where the cycle no longer satisfies the conditions,
\begin{equation}
	\langle Q_2 \rangle > 0, \quad \langle Q_4 \rangle < 0, \quad \text{and}\quad \langle W_1 \rangle + \langle W_3 \rangle = \langle W_{\mathrm{tot}} \rangle < 0, 
\end{equation}
where we use the convention that work or heat flowing into the system is positive.
These are known as the \textit{positive work conditions} and must be met for the cycle to function as an engine. In the parameter regimes where the positive work conditions are not satisfied, efficiency and power are no longer meaningful metrics of performance. In these regimes, the cycle functions instead as a heater, accelerator, or refrigerator. In a heater, work is put into the system to induce heat flow into both baths, ($\langle Q_2 \rangle < 0, \langle Q_4 \rangle < 0, \langle W_{\mathrm{tot}} \rangle > 0$). In an accelerator, work is put into the system to enhance the flow of heat from the hot to the cold bath ($\langle Q_2 \rangle > 0, \langle Q_4 \rangle < 0, \langle W_{\mathrm{tot}} \rangle > 0$). Lastly, in a refrigerator, work is put into the system to induce heat flow from the cold to the hot bath ($\langle Q_2 \rangle < 0, \langle Q_4 \rangle > 0, \langle W_{\mathrm{tot}} \rangle > 0$). 

\begin{figure}[H]
\begin{minipage}[c]{.3\textwidth}
	\centering
	\includegraphics[width=1.05\textwidth]{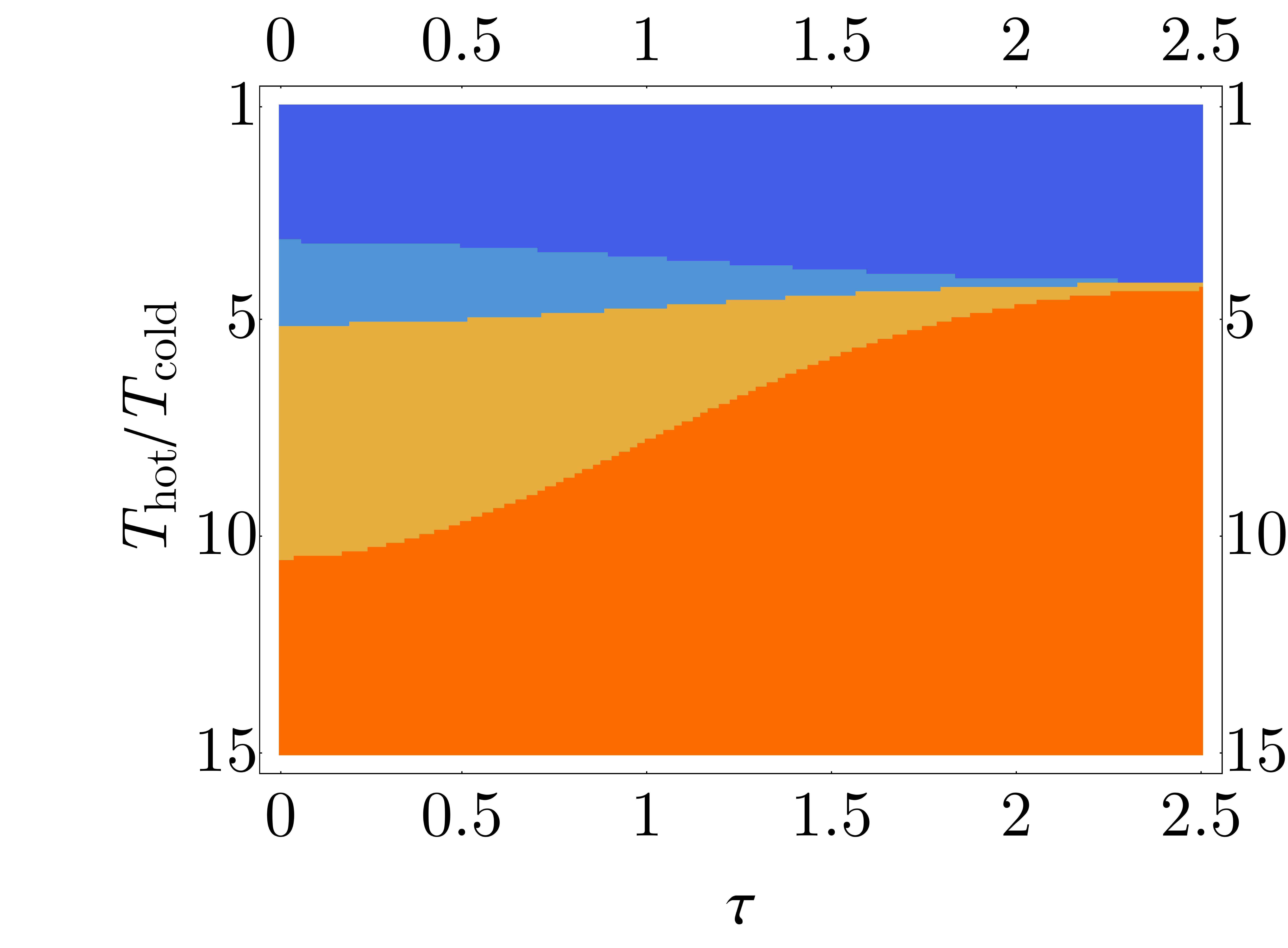}
	\caption*{\qquad\qquad\qquad\qquad\qquad (\textbf{a})}
\end{minipage}%
\begin{minipage}[c]{0.3\textwidth}
	\centering
	\includegraphics[width=1.05\textwidth]{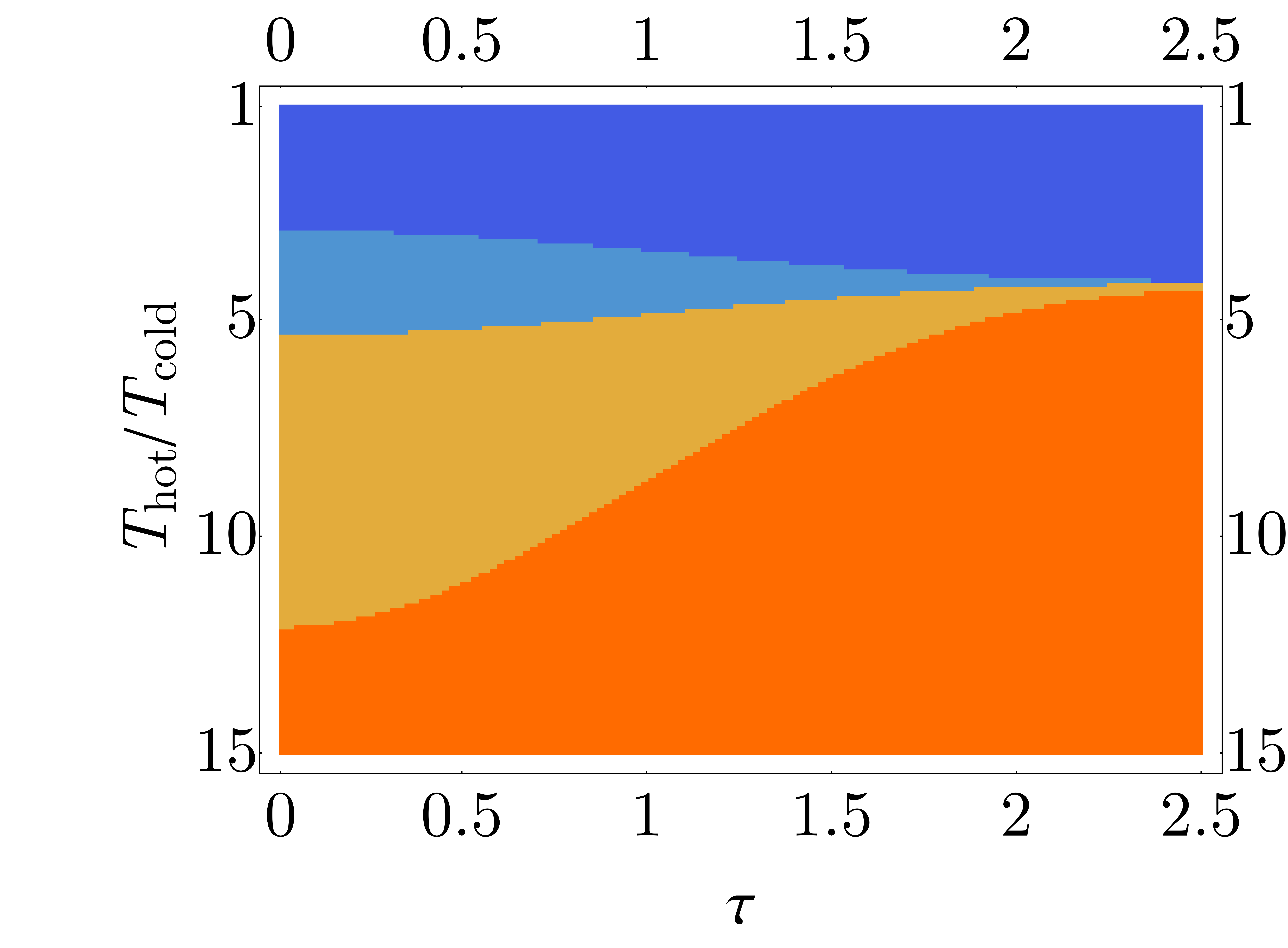}
	\caption*{\qquad\qquad\qquad\qquad\qquad (\textbf{b})}
\end{minipage}
\begin{minipage}[c]{0.3\textwidth}
	\centering
	\includegraphics[width=1.05\textwidth]{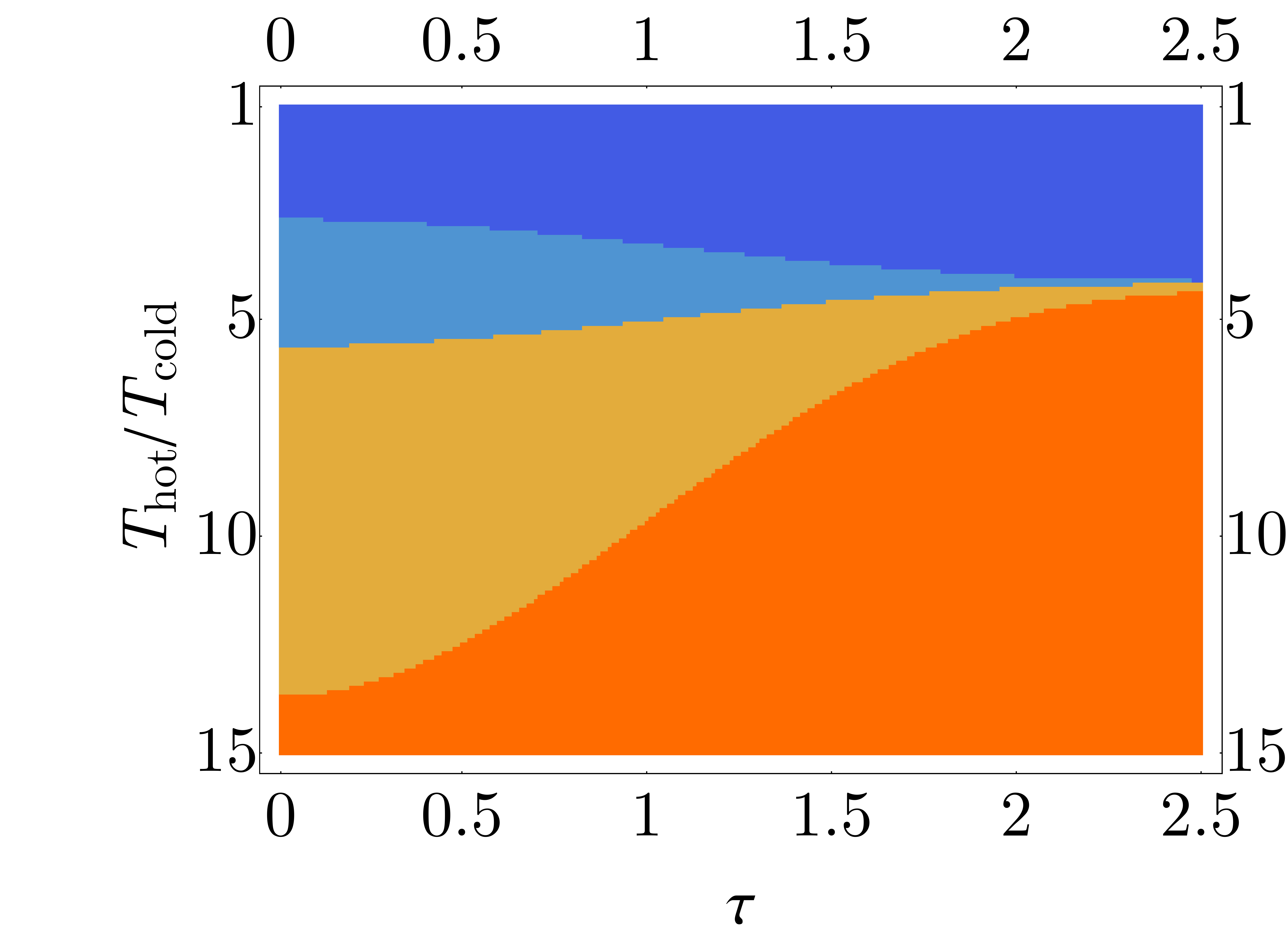}
	\caption*{\qquad\qquad\qquad\qquad\qquad (\textbf{c})}
\end{minipage}
\caption{\label{ParamReg} Parameter regimes where the cycle functions as a refrigerator (dark blue, top), heater (light blue, upper middle), accelerator (tan, lower middle) and engine (orange, bottom) for (a) $\nu = 0$, (b) $\nu = 0.5$, and (c) $\nu = 1$. Parameters are $\omega_0 = \delta\omega = 1$.}
\end{figure}

We note that the values of the bath temperature ratio and stroke time for which the system transitions out of the engine regime vary for different values of $\nu$. To further explore this relationship, in Figure~\ref{ParamReg}, we show the parameter space under which the cycle functions as each type of thermal machine for different interaction strengths. We see that for short stroke times, $\nu = 0$ displays the smallest heater and accelerator regimes, and the largest engine regime, while $\nu =0$ displays the largest heater and accelerator regimes, and the smallest engine regime. Intermediate values of $\nu$ fall between these limits. In general, we see that as cycle time increases, the heater and accelerator regimes vanish, as do the differences in the parameter space for the different values of $\nu$.    

\end{paracol}
\nointerlineskip
\begin{paracol}{2}
	\switchcolumn   
    
\vspace{-9pt}

%%%%%%%%%%%%%%%%%%%%%%%%%%%%%%%%%%%%%%%%%%
\section{Discussion}

In this work, we have analyzed the performance of a quantum Otto engine with a working medium of two {interacting} particles in a singular oscillator potential, including the efficiency, power output, and parameter regimes where the cycle functions as different types of thermal machines. We have determined the full time-dependent dynamics using the framework of statistical anyons and shown that by changing the strength of the interparticle interaction, we can interpolate between the performance of bosonic and fermionic working mediums. We see that the impact of the singular interaction on the engine performance vanishes for long stroke times, as the engine approaches fully adiabatic performance, indicating that the performance differences are a fundamentally nonequilibrium phenomena. {Consistent with the results of \cite{Myers2020}, we have found that the best performance in regards to both efficiency and power arises in the limit of $\nu = 0$, in which the interparticle interaction vanishes and the particles behave as ideal bosons in a pure harmonic potential. While interacting particles yield reduced performance in comparison to ideal bosons, they yield enhanced performance in comparison to ideal fermions. As performance depends directly on the interaction, such an engine may have other applications as well, for instance using differences in performance to measure the generalized exclusion statistics parameter of an anyonic system.} 
	
{By incorporating interparticle interactions, we extend our exploration of the impact of wave function symmetry on heat engine performance to a more physically realistic model. The inverse square interaction potential in particular arises in the case of electron--dipole interactions \cite{LevyLeblond1967, Jaramillo2010} and, as previously noted, as the basis of the Calogero--Sutherland model. In general, solving the dynamics of interacting systems is significantly more challenging than in the idealized case. Here, we demonstrate the usefulness of the statistical anyon framework in overcoming this challenge, as it provides a simple and novel method for mapping the dynamics of interacting particles to that of a mixture of ideal bosons and fermions.}      

Introducing a time-dependent interaction strength would open up the possibility of optimizing performance by varying the behavior of the particles between that of bosons and fermions during the engine cycle. {Engine power output could also be optimized by minimizing the duration of the compression and expansion strokes while suppressing nonadiabatic excitations \cite{Stefanatos2017}, which can be accomplished using shortcuts to \mbox{adiabaticity~\cite{Torrontegui2013, GueryOdelin2019}}. However, for an accurate performance assessment in such cases, it is important that the cost of implementing the shortcut is also accounted for \cite{Campbell2017, Abah2017, Abah2018, Baris2019, Abah2019, Torrontegui2017, Tobalina2019}.} We leave these as topics to be explored in future work.    
	
The Calogero--Sutherland model described by Equation (\ref{2partHamil}) has received extensive study in the context of generalized exclusion statistics, and due to its direct connection to spin-chain models, such as the Haldane--Shastry chain \cite{Haldane1988, Shastry1988, Polychronakos1993, Haldane1994}, which provide a promising route for experimental realization in trapped atom systems \cite{Grass2014, Britton2012, Bohnet2016, Labuhn2016}. Such systems may provide a means of experimentally implementing a singular quantum heat engine. Furthermore, generalized exclusion statistics anyons can be used to replicate the thermodynamic behavior of fractional exchange statistics anyons that manifest in two-dimensional systems~\cite{Myers2021}. These anyons are highly sought after due to their close relationship to non-ableian anyons, which may be used to implement fault-tolerant topological quantum computers \cite{Moore1991}. {However, direct observation and manipulation of fractional exchange statistics anyons is extremely difficult \cite{Bartolomei2020}.} Using the well-established framework of heat engines to understand the thermodynamic behavior of systems that display intermediate statistics, {such as the singular oscillator examined here,} may lead to better methods of detection and control for fractional exchange statistics anyons.    
 
%%%%%%%%%%%%%%%%%%%%%%%%%%%%%%%%%%%%%%%%%%
\vspace{6pt} 

%%%%%%%%%%%%%%%%%%%%%%%%%%%%%%%%%%%%%%%%%%
%% optional
%\supplementary{The following are available online at \linksupplementary{s1}, Figure S1: title, Table S1: title, Video S1: title.}


%%%%%%%%%%%%%%%%%%%%%%%%%%%%%%%%%%%%%%%%%%
\authorcontributions{Conceptualization, S.D.; methodology, N.M.M. and S.D.; formal analysis, N.M. and J.M.; writing---original draft preparation, N.M.M.;  writing---review and editing, N.M.M. and S.D.; supervision, S.D.; funding acquisition, S.D. All authors have read and agreed to the published version of the manuscript.}

\funding{S.D. acknowledges support from the U.S. National Science Foundation under Grant No. DMR-2010127.}

\institutionalreview{Not applicable %MDPI: .In this section, please add the Institutional Review Board Statement and approval number for studies involving humans or animals. Please note that the Editorial Office might ask you for further information. Please add ``The study was conducted according to the guidelines of the Declaration of Helsinki, and approved by the Institutional Review Board (or Ethics Committee) of NAME OF INSTITUTE (protocol code XXX and date of approval).'' OR ``Ethical review and approval were waived for this study, due to REASON (please provide a detailed justification).'' OR ``Not applicable'' for studies not involving humans or animals. You might also choose to exclude this statement if the study did not involve humans or animals.
}

\informedconsent{Not applicable %MDPI: .Any research article describing a study involving humans should contain this statement. Please add ``Informed consent was obtained from all subjects involved in the study.'' OR ``Patient consent was waived due to REASON (please provide a detailed justification).'' OR ``Not applicable'' for studies not involving humans. You might also choose to exclude this statement if the study did not involve humans.

%Written informed consent for publication must be obtained from participating patients who can be identified (including by the patients themselves). Please state ``Written informed consent has been obtained from the patient(s) to publish this paper'' if applicable.
}

\dataavailability{Not applicable %MDPI: .In this section, please provide details regarding where data supporting reported results can be found, including links to publicly archived datasets analyzed or generated during the study. Please refer to suggested Data Availability Statements in section ``MDPI Research Data Policies'' at \url{https://www.mdpi.com/ethics}. You might choose to exclude this statement if the study did not report any data.
}

\acknowledgments{This work was conducted as part of the Undergraduate Research Program (J.M.) in the Department of Physics at UMBC.}

\conflictsofinterest{The authors declare no conflict of interest.} 

%%%%%%%%%%%%%%%%%%%%%%%%%%%%%%%%%%%%%%%%%%

\pagebreak

\appendixtitles{no} % Leave argument "no" if all appendix headings stay EMPTY (then no dot is printed after "Appendix A"). If the appendix sections contain a heading then change the argument to "yes".
\appendixstart
\appendix
\section{}
\label{AppendixA}
In this appendix we provide the full expressions for the time-dependent density operators for two bosons and two fermions in a harmonic potential. The density operator for bosons is,

\vspace{-9pt}

\begin{adjustwidth}{-4.6cm}{0cm}
\begin{equation}
	\begin{split}
		&\rho_{\mathrm{har}}^{(\mathrm{B})}(x_1,x_2,y_1,y_2;t)= \frac{m \omega}{2 \pi \hbar (W_t^2+Z_t^2 \omega^2)} \left(e^{-\beta \hbar \omega}-1\right) \\
		&\quad\times\bigg\{ e^{\frac{m}{2 \hbar (W_t^2+Z_t^2 \omega^2)}\left[i (x_1^2+x_2^2-y_1^2-y_2^2)(W_t \dot{W}_t+Z_t\dot{Z}_t\omega^2)-\omega (x_1^2+x_2^2+y_1^2+y_2^2) \mathrm{coth}(\beta \hbar \omega)
			+2 \omega (x_1 y_1 +x_2 y_2) \mathrm{csch}(\beta \hbar \omega) \right]} \\
		&\quad+ e^{\frac{m}{2 \hbar (W_t^2+Z_t^2 \omega^2)}\left[i (x_1^2+x_2^2-y_1^2-y_2^2)(W_t \dot{W}_t+Z_t\dot{Z}_t\omega^2)
			- \omega(x_1^2+x_2^2+y_1^2+y_2^2) \mathrm{coth}(\beta \hbar \omega)+2 \omega (x_2 y_1 +x_1 y_2) \mathrm{csch}(\beta \hbar \omega) \right]} \bigg\}.
	\end{split}
\end{equation}
\end{adjustwidth}

The density operator for fermions is,

\vspace{-9pt}
\begin{adjustwidth}{-4.6cm}{0cm}
\begin{equation}
	\begin{split}
		&\rho_{\mathrm{har}}^{(\mathrm{F})}(x_1,x_2,y_1,y_2;t)= \frac{m \omega}{2 \pi \hbar (W_t^2+Z_t^2 \omega^2)} \left(e^{ \beta \hbar \omega}-1\right) \\
		&\quad\times\bigg\{ e^{\frac{m}{2 \hbar (W_t^2+Z_t^2 \omega^2)}\left[i (x_1^2+x_2^2-y_1^2-y_2^2)(W_t \dot{W}_t+Z_t\dot{Z}_t\omega^2)-\omega (x_1^2+x_2^2+y_1^2+y_2^2) \mathrm{coth}(\beta \hbar \omega)
			+2 \omega (x_1 y_1 +x_2 y_2) \mathrm{csch}(\beta \hbar \omega) \right]} \\
		&\quad- e^{\frac{m}{2 \hbar (W_t^2+Z_t^2 \omega^2)}\left[i (x_1^2+x_2^2-y_1^2-y_2^2)(W_t \dot{W}_t+Z_t\dot{Z}_t\omega^2)
			- \omega(x_1^2+x_2^2+y_1^2+y_2^2) \mathrm{coth}(\beta \hbar \omega)+2 \omega (x_2 y_1 +x_1 y_2) \mathrm{csch}(\beta \hbar \omega) \right]} \bigg\}.
	\end{split}
\end{equation}
\end{adjustwidth}

Note that $Z_t$ and $W_t$ are the same as in Equation~(\ref{ClassicalEOM}).

\section{}
\label{AppendixB} 

In this appendix we provide the full expressions for the engine efficiency and power output. The efficiency is,
\begin{adjustwidth}{-4.6cm}{0cm}
\begin{equation}
	\eta = 1 + \frac{\omega_0}{\omega _{\tau }}\left[\frac{Q^*_{3} \left(-3 \coth \left(\beta _h \hbar \omega_{\tau}\right)-\text{csch}\left(\beta _h \hbar \omega_{\tau}\right)+2 \nu -1\right)+3 \coth \left(\beta_c \hbar \omega _0 \right)+\text{csch}\left(\beta_c \hbar \omega _0\right)-2 \nu +1}{Q^*_{1} \left(-3 \coth \left(\beta_c \hbar \omega _0\right)-\text{csch}\left(\beta_c \hbar \omega _0\right)+2 \nu -1\right)+3 \coth \left(\beta _h \hbar \omega_{\tau} \right)+\text{csch}\left(\beta _h \hbar \omega_{\tau}\right)-2 \nu+1}\right],
\end{equation}
\end{adjustwidth}
where $Q^*_1$ is the adiabaticity parameter for stroke 1 (compression) and $Q^*_3$ is the adiabaticity parameter for stroke 3 (expansion). The power output is,

\vspace{-12pt}
\begin{adjustwidth}{-4.6cm}{0cm}
\begin{equation}
	\begin{split}
	P = & \frac{\hbar}{4 \gamma  \tau }  \big[\omega _0 \left(3 \coth \left(\beta_c \hbar \omega_0 \right)+\text{csch}\left(\beta_c \hbar \omega_0\right)-2 \nu +1\right)+\omega _{\tau } (Q^*_1 \left(-3 \coth\left(\beta_c \hbar \omega_0\right)-\text{csch}\left(\beta_c \hbar \omega_0\right)+2 \nu -1\right)
	\\ &+3 \coth \left(\beta_h \hbar \omega_{\tau} \right)+\text{csch}\left(\beta_h \hbar \omega_{\tau} \right)-2 \nu+1)+Q^*_3 \omega _0 \left(-3 \coth \left(\beta_h \hbar \omega_{\tau} \right)-\text{csch}\left(\beta_h \hbar \omega_{\tau} \right)+2 \nu-1\right)\big].
	\end{split}
\end{equation}
\end{adjustwidth}

 \vspace{-12pt}

%%%%%%%%%%%%%%%%%%%%%%%%%%%%%%%%%%%%%%%%%%
\end{paracol}
%%%%%%%%%%%%%%%%%%%%%%%%%%%%%%%%%%%%%%%%%%
\reftitle{References}

\end{document}